\documentstyle[art8, aas2pp4, 10pt]{article}
\def\taud{\tau}
\def\simless{\lesssim}
\def\simgreat{\gtrsim}

\def\dgc{{D_{\rm GC}}}
\def\deltagc{{\Delta_{\rm GC}}}

\begin{document}

\title{\huge FINDING RADIO PULSARS IN AND BEYOND THE GALACTIC CENTER}
\author{\huge James M. Cordes \& T. Joseph W. Lazio}
\affil{\huge\rm Dept. of Astronomy and NAIC}
\affil{\Large\rm Cornell University, Ithaca, NY  14853-6801 USA} 
\affil{\Large \rm cordes@spacenet.tn.cornell.edu}
\affil{\Large \rm lazio@spacenet.tn.cornell.edu}

\vskip 0.1truein
\centerline{submitted to {\it The Astrophysical Journal}}
\centerline{1996 May}

\lefthead{Cordes \& Lazio}
\righthead{Galactic Center Pulsars}

\begin{abstract}
\huge
Radio-wave scattering is enhanced dramatically for galactic center
sources in a region with radius $\simgreat 15$~arc~min.  Using
scattering from Sgr~A${}^*$ and other sources, we show that pulse
broadening for pulsars in the Galactic center is {\em at least\/}
$6.3\,\nu^{-4}$~seconds ($\nu$ = radio frequency in GHz) and is most
likely 50--200 times larger because the relevant scattering screen
appears to be within the Galactic center region itself.  Pulsars
beyond---but viewed through---the Galactic center suffer even
greater pulse broadening and are angularly broadened by up to $\sim 2$
{\em arc~min}.  Periodicity searches at radio frequencies are likely to find
only long period pulsars and, then, only if optimized by using
frequencies $\simgreat 7$~GHz and by testing for small numbers of
harmonics in the power spectrum.  The optimal frequency is $\nu \sim
7.3\,{\rm GHz} (\Delta_{0.1}P\sqrt{\alpha})^{-1/4}$ where
$\Delta_{0.1}$ is the distance of the scattering region from Sgr~A${}^*$ in
units of 0.1~kpc, P is the period (seconds), and $\alpha$ is the
spectral index.  A search for compact sources using aperture synthesis
may be far more successful than searches for periodicities because the
angular broadening is not so large as to desensitize the survey.  We
estimate that the number of {\em detectable\/} pulsars in the Galactic
center may range from $\le 1$ to 100, with the larger values 
resulting from recent, vigorous starbursts.  Such pulsars provide unique
opportunities for probing the ionized gas, gravitational potential,
and stellar population 
near Sgr~A${}^*$.
\end{abstract}

\keywords{scattering --- pulsars: general --- Galaxy: center}

\section{INTRODUCTION}\label{sec:intro}

Only 15 radio pulsars of the current catalog of approximately 700 are
within 5\arcdeg\ of Sgr~A$^*$ and none are within 1\arcdeg\ (Taylor,
Manchester, \& Lyne~1993).  Discovering pulsars in or near the
Galactic center (GC) would provide exciting opportunities for probing
magnetoionic material, the gravitational potential, and star formation
in the GC (e.g., \cite{har95}; \cite{sof94}).  Counterparts to X- and
$\gamma$-ray GC sources may also involve radio pulsars or their
progenitors.  Moreover, dynamical constraints on the mass in the GC
combined with infrared observations suggest that neutron stars may
abound there.

It may be impossible to find radio pulsars in the GC as sources of
periodic radio emission, however.  The difficulty arises because
radio-wave scattering, which is particularly severe from sources in
the GC (e.g., Sgr~A${}^*$, \cite{b+93}; \cite{lo+93}; \cite{b88}; and
several OH/IR stars, \cite{fdcl94}; \cite{van92}), broadens pulsar
pulses by extraordinary amounts.  The actual level of angular
broadening toward the GC at 1~GHz is $\sim 10$ times greater than that
predicted by a recent model for the distribution of free electrons in
the Galaxy (\cite{tc93}; hereafter TC), even though this model
includes a general enhancement of scattering toward the inner Galaxy.
For pulsars in the GC, the resultant pulse broadening is at least 100
times larger than predicted by the TC model.  The TC model purposely
excluded an explicit treatment of the electron density in the GC,
primarily because it would have been, at best, poorly quantified.
More recent data on angular broadening of GC sources, however, now
allows us to estimate the amount by which pulse broadening should be
augmented.  Our results complement recent work (\cite{j94}) that 
suggests a paucity of pulsars inside of the molecular
ring at Galactocentric radius $\sim 4$~kpc.   Johnston~(1994) considered
a large-scale region toward the inner Galaxy and did not discuss
the immediate vicinity of the GC.

Pulse broadening is so large in the GC that, at conventional frequencies used
for pulsar searches, 0.4--1.4~GHz, the pulsed flux is orders of
magnitude too small to be detected, as we show below.  The strong
frequency dependence of pulse broadening ($\propto\nu^{-4}$) suggests
that frequencies at much higher frequencies, 5--20~GHz, will mitigate
the effects of scattering.  But pulsar spectra generally decline at
higher frequencies, sometimes precipitously, making this option
unlikely to find many pulsars.  However, some pulsars have been
detected at frequencies as high as 30~GHz (\cite{xil+95};
\cite{seir+95}; \cite{izve+94}; \cite{kramer+94}; \cite{mal94}).
Though these are relatively nearby objects, their detections suggest
that some pulsars may show sufficient spectral flattening that they
would be detectable if as far as the GC.

In this paper we discuss pulse broadening from sources in and beyond
the GC. We show how to optimize a periodicity search in terms of radio
frequency and number of harmonics.  We also discuss searches for
pulsars as steady sources rather than periodic ones.  Namely, we show
that aperture synthesis surveys using the Very Large Array (VLA), for
example, should be successful in finding objects with pulsar-like
spectra and polarization.

In \S\ref{sec:geometry} we detail the assumed geometry and angular
broadening.  In \S\ref{sec:pulsebroad} we derive the resulting pulse
broadening timescales. \S\ref{sec:detection} discusses the pessimistic
implications of pulse broadening for pulsar surveys involving
periodicity searches.  In \S\ref{sec:imaging} we argue that aperture
synthesis surveys may be more successful than periodicity searches in
identifying radio pulsars in the GC.  In \S\ref{sec:population} we
estimate the likely number of pulsars in the GC.  In
\S\ref{sec:conclude} we present our conclusions and recommendations.

\section{ANGULAR BROADENING AND THE SCATTERING GEOMETRY}\label{sec:geometry}

The angular diameter of Sgr~A$^*$ is 1.3\arcsec\ at 1~GHz and scales
as $\nu^{-2}$, in accord with interstellar scattering (\cite{lo+93}). 
OH/IR stars in the vicinity of the GC also show very heavy scattering,
at least for those stars within $\sim 15\arcmin$ of Sgr~A$^*$
(\cite{fdcl94}; \cite{van92}).  We use the measured angular
broadening to predict the pulse broadening expected from pulsars in
the GC, but with the following caveat: we must assume the location
along the line of sight of the scattering ``screen'' that enhances
the scattering.  As van~Langevelde et al.~(1992) discuss, the location
of the screen is by no means determined and has implications for the
depth of free-free absorption to the GC and the outer scale of the
electron-density variations.  For now, we treat the location of the
screen as a parameter of the overall geometry.  We constrain its
location using recent scattering measurements
for the GC region (Lazio, Cordes \& Frail~1996).

\begin{figure}[tb]
\plotfiddle{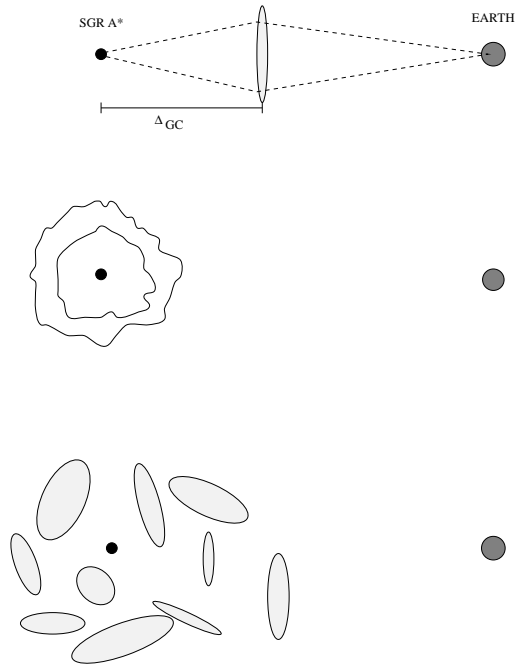}{4.0truein}{0}{45}{45}{-144}{-50}
\caption[Scattering Geometry]
{Geometries for the Galactic Center scattering region.
(Top) A single, thin screen located a distance $\deltagc$ from the
Galactic center.
This geometry is used in most of the dicusssion in this paper. 
(Middle) A spherical (or cylindrical) shell surrounding the GC.
(Bottom) A scattering region composed of ``screenlets'' distributed
around the GC with total size $\sim \deltagc$.} 
\label{fig:geometry}
\end{figure}

We model the scattering as arising from a discrete screen, as in the
top of Fig.~\ref{fig:geometry}.  Let $\theta_{\rm s}$ be the
scattering angle produced by the screen and $\theta_{\rm o}$ the
observed scattering angle.  In the following we assume the small-angle
approximation: $\theta_{\rm s}, \theta_{\rm o} \ll 1$.  For a source
at infinite distance, $\theta_{\rm o} \equiv \theta_{\rm s}$ because
the waves incident on the screen are plane waves.  Sources at finite
distance are scattered less due to wave sphericity.  If the source is
at distance $D$ and the screen is in front of the source by distance
$\Delta$, the observed scattering angle is 
$\theta_{\rm o} = \left({\Delta}/{D} \right)\theta_{\rm s}$.  
If the scattering screen is
very close to the GC, then $\Delta \ll D$ and the observed scattering
diameter of an extragalactic source is
\begin{equation}
\theta_{\rm xgal} = \left( \frac{\dgc}{\deltagc} \right)\theta_{\rm GC} \gg 
\theta_{\rm GC},
\label{eq:xgal}
\end{equation}
where $\theta_{\rm GC}$ is the observed scattering diameter of
Sgr~A${}^*$.  If the scattering screen is, in fact, associated with
the GC (e.g., $\dgc \simless 100$~pc), then extragalactic sources
viewed through the screen necessarily have angular diameters that
exceed 1~{\em arc~min\/} at 1~GHz.

The screen geometry we have utilized is highly idealized.  Other, more
physical, geometries are also presented in Fig.~\ref{fig:geometry}.
The scattering screen may {\em surround\/} the GC with spherical or
cylindrical geometry.  Alternately, the scattering region may be
distributed throughout the GC region rather than being confined to a
thin screen.  Yusef-Zadeh et al.~(1994), for example, interpret
anisotropic scattering from Sgr~A${}^*$ in terms of \ion{H}{2} regions
distributed uniformly within 100~pc of Sgr~A${}^*$.  The scattering
{}from Sgr~A${}^*$ may then be {\em effectively\/} screen-like because
it will be dominated by \ion{H}{2} regions farthest from Sgr~A${}^*$
(through geometric effects).  However, background sources will be
scattered by all \ion{H}{2} regions.  In these cases, the angular
diameters of background sources are at least double what we have
derived if the scattering is from a thin screen that surrounds the GC,
but it will be even greater for scattering regions that fill the GC.
 
Van~Langevelde et al.~(1992) derived a lower bound on the GC-screen
distance based on the fact that the free-free optical depth must be
$\simless 1$ at 1.6~GHz:
\begin{eqnarray}
\lefteqn{\deltagc \simgreat} \nonumber \\ 
 && 0.1 \dgc \left(\frac{\ell_1}{100\,{\rm km}} \right )^{1/6}
       \left(\frac{\ell_0}{1\,{\rm pc}} \right )^{1/3}
       \left[ \frac{g(\nu,T)} {T_4^{1.5}} \right]^{1/2}.
\label{eq:freefree}
\end{eqnarray}
In this equation, $\ell_{0}$, $\ell_1$ are the outer and inner scales,
respectively, of the electron-density fluctuations; $g(\nu,T)$ is the
Gaunt factor, and $T_4$ is the temperature in units of $10^4$~K.  The
various parameters can be adjusted to reduce the overall 
coefficient from 0.1 to~0.004 (e.g., through an inner scale of 0.01~pc and a
temperature of $10^5$~K).  However, even $10^5$~K may be too hot for
electron density variations to be sustained (e.g., \cite{s91}).  A
scattering screen thinner than 0.01~pc is also improbable because
stellar-wind shocks tend to be at least this thickness (\cite{kb90}).
The outer scale, which is probably comparable to the screen thickness,
cannot be reduced much below 0.01~pc.  Other evidence also indicates
that the inner scale is not less than 100~km.  This includes the
observed scaling of scattering diameters as $\nu^{-2}$ for Sgr~A${}^*$
and other heavily scattered sources such as Cyg X-3 (\cite{mmrj95};
Wilkinson, Narayan, \& Spencer~1994) and NGC6334B (\cite{mgrb90}).  We
therefore take as a hard lower bound, $\deltagc \ge 33$~pc.  This
suggests that a strong upper bound on the sizes of extragalactic
sources is $\theta_{\rm xgal} \simless 11\arcmin$ for a screen-like
geometry that wraps around Sgr~A${}^*$.

It is possible to determine the screen distance from the Galactic
center by either measuring the diameters of extragalactic sources or,
in the case where severe scattering diminishes our ability to detect
sources, use the deficiency of sources to constrain the screen
location.  Attempts to observe extragalactic sources near Sgr~A${}^*$
in fact show a deficiency of sources within 0.5\arcdeg\ of Sgr~A${}^*$
(\cite{lcf96}).  A likelihood analysis shows that this deficiency is
consistent with extragalactic source sizes larger than 0.5\arcmin\ and
$\deltagc \lesssim 200$~pc (\cite{lcf96}).

\section{PULSE BROADENING}\label{sec:pulsebroad}

Pulse broadening is a diffraction phenomenon but may be treated with
stochastic ray tracing through extended media (Williamson~1972, 1975).
It has been observed for many pulsars and used to study the
distribution of ionized microturbulence in the Galaxy (\cite{c+91}).

The pulse broadening due to the screen responsible for the scattering
of Sgr~A${}^*$ has an $e^{-1}$ time scale
\begin{equation}
\taud_{\rm GC}(\dgc) \sim 
  f\left( \frac{\deltagc}{\dgc} \right )
  \left ( \frac{\dgc\theta_{\rm GC}^2}{8c\ln 2} \right ),
\end{equation}
where the screen's location along the line of sight is represented by
the geometric factor\footnote{Pulse
broadening is often expressed in terms of the {\em screen\/} scattering
angle $\theta_{\rm s}$ rather than the observed angle $\theta_{\rm o}$.  Using
$\theta_{\rm s}$, the equivalent geometric factor is $x(1-x)$, which maximizes
at $x=1/2$. }
\begin{equation}
f(x) \equiv x^{-1}(1-x).
\end{equation}
The pulse broadening time for fiducial values of the distance
(8.5~kpc) and scattering diameter (1.3\arcsec) at a frequency of 1~GHz
is
\begin{eqnarray}
\lefteqn{\taud_{\rm GC}(\dgc) \sim } \nonumber \\ 
 && 6.\!^{\rm s}3\,\left(\frac{\dgc}{8.5\,\rm kpc}\right)
    \left(\frac{\theta_{\rm GC, 1 \,GHz}} {1.3\arcsec}\right)^2
     \nu_{\rm GHz}^{-4} f\left(\frac{\deltagc} {\dgc}\right).
\label{eq:taugc}
\end{eqnarray}
In Eq.~\ref{eq:taugc} we have adopted a frequency scaling
$\propto \nu^{-4}$ rather than the often encountered $\nu^{-4.4}$ scaling
because, in the extremely strong scattering limit, the scattering is dominated
by the smallest irregularities in the free electron density
that are physically present (cf.\ \cite{cl91}).  This is consistent with
the observed $\nu^{-2}$ scaling of the angular diameter of Sgr~A${}^*$. 
The geometric factor is $f \to 1$ if the screen is midway along the
line of sight.  But for screens very near the GC, $f\to x^{-1} \gg 1$.
Therefore, pulsars at the same location as the GC will show {\em at
least\/} $6.3$~s of pulse broadening at 1~GHz\footnote{
An early analysis (Davies, Walsh \& Booth~1976) estimated 10~s of
pulse broadening at 1~GHz while implicitly assuming the scattering
region to be midway between us and the GC.
}.  
The pulse broadening may be significantly larger, perhaps as much as
200 times larger, because the scattering screen may be only 33--100~pc
{}from the GC.  The minimal scattering time of $6.3$~s may be compared,
at 1~GHz, to the pulse broadening of the most heavily scattered
pulsar, PSR~B1849$-$00 (\cite{fc89}; \cite{clif+92}), which is about
0.3~s.

Pulsars beyond the GC (but still behind the
scattering screen) will show even  larger scattering.  
For a pulsar distance $D \ge \dgc - \deltagc$, the pulsar-screen distance
is $\Delta \equiv  D - \dgc + \deltagc$ and the 
pulse broadening from the screen is 
\begin{equation}
\taud_{\rm GC}(D) \sim \taud_{\rm GC}(\dgc)
                        \left ( \frac{\dgc} {D} \right )
                        \left ( \frac{\Delta}{\deltagc} \right ).
\end{equation}
As a function of distance from the Sun, 
pulse broadening increases slowly and according to the TC
model, which possesses components that grow stronger in the inner
Galaxy.  Then, just beyond the location of the GC scattering screen,
pulse broadening increases dramatically and continues to increase.
To combine the TC model and the GC screen component, we write the
net pulse broadening as 
\begin{equation}
 \taud =  \cases {
     \taud_{\rm TC},  &  $D < \dgc - \deltagc$;\cr 
\cr
     ( \taud_{\rm TC}^2 + \taud_{\rm GC}^2 )^{1/2}, & $D \ge \dgc - \deltagc$.
}
\label{eq:tau_total}
\end{equation} 
Combining the TC and GC-screen scattering times is {\em ad hoc\/}
in form but is sufficiently accurate for our purposes here because
the GC component is much larger than the TC contribution. 

\begin{figure}[tbh]
\plotfiddle{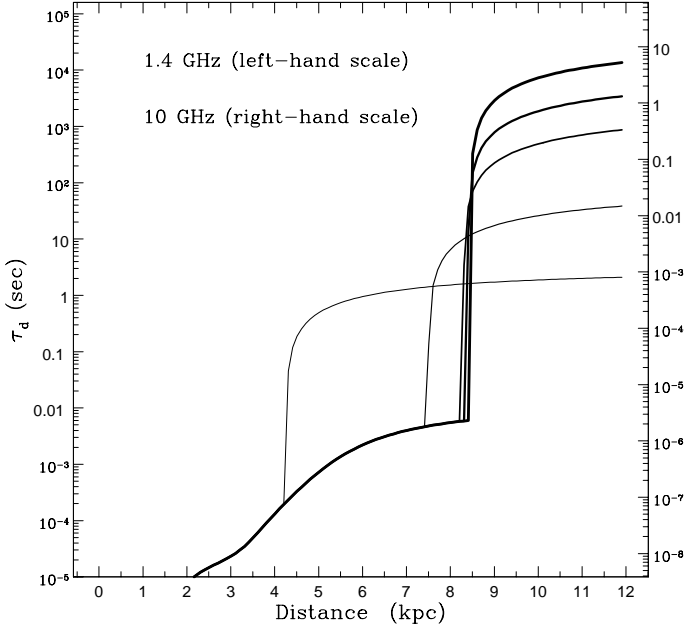}{4.0truein}{0.}{45}{45}{-144}{-52}
\caption[Pulse Broadening]{
Pulse broadening is plotted against distance for 5 separate
values of the GC-screen distance, $\deltagc$:
0.05~kpc (thickest line), 0.1, 0.2, 1.0, and 4.25~kpc (thinnest line).
The left-hand scale applies to 1.4~GHz, the right-hand scale to 10~GHz.
Broadening at other frequencies may be estimated using the assumed
$\nu^{-4}$ scaling.
}
\label{fig:gctau}
\end{figure}

Figure~\ref{fig:gctau} shows the pulse broadening at two frequencies
(1.4 and 10~GHz) for a range of GC-screen distances, $\deltagc =
0.05$, 0.1, 0.2, 1.0, and 4.25~kpc.  For pulsars beyond the GC, the
pulse broadening asymptotes to $\taud_{\rm GC}(\dgc)(\dgc/\deltagc)
\sim 10^5$~s at 1.4~GHz and 18~s at 10~GHz.

\section{DETECTION OF SCATTERED PULSARS IN PERIODICITY SEARCHES}
\label{sec:detection}

Pulse broadening decreases the number of harmonics that exceed
a predetermined threshold in the power spectrum of the intensity,
thus reducing the sensitivity of a pulsar search. 
Consider a train of pulses with period $P$, average pulse area $A_0$,
duty cycle $\epsilon$, and pulse
width (FWHM) $W\equiv\epsilon P$.   
The discrete Fourier transform of the pulse train is a
series of spikes at frequencies $\ell/P, \ell = 0, 1, \ldots$ each
having an amplitude,
\begin{equation}
{\cal A}_{\rm DFT}(\ell)
 = A_0\tilde g(\epsilon \ell),
\label{eqn:dftsnr}
\end{equation}
where pulses have a generic shape $g(\phi)$ in pulse phase $\phi$, 
whose continuous Fourier transform is $\tilde g$. 

We define the {\em intrinsic\/} pulsed fraction of the pulsar flux as the ratio of 
the fundamental frequency and zero frequency (``DC'') amplitudes:
\begin{equation}
\eta_{P} \equiv
\left\vert\frac{ {\cal A}_{\rm DFT}(1) }
     { {\cal A}_{\rm DFT}(0)} \right\vert
 = \left\vert\frac{\tilde g(\epsilon)}{\tilde g(0)}\right\vert
 = \exp \left[ - \left ( \frac{\pi\epsilon} {2\sqrt{\ln 2} } \right )^2 \right].
\label{eq:pfraction1}
\end{equation}
where the third equality is for gaussian-shaped pulses [i.e., $g(\phi)
= \exp(-4\ln 2\phi^2)$].  For most pulsars, $\epsilon \simless 0.1$
implying $\eta_P \sim 1$.

\begin{figure}[tbh]
\plotfiddle{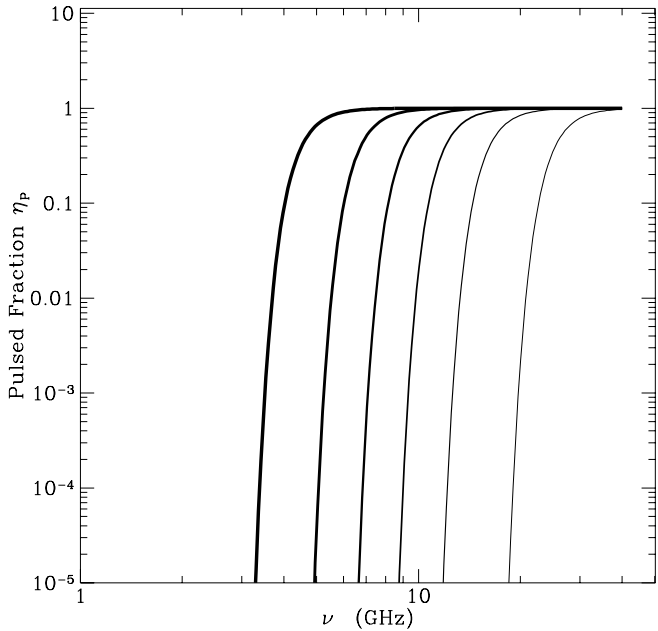}{4.0truein}{0.}{45}{45}{-144}{-52}
\caption[Pulsed Fraction]{
Plot of the pulsed fraction $\eta_{P}^{\rm (s)}$ 
against frequency $\nu$ for different pulse periods,  
$P = 5$~s  (thickest line), 1~s, 0.3~s, 0.1~s, 30~ms, \& 5~ms (thinnest line).
The scattering screen is assumed to be close to the GC,
$\deltagc = 50$~pc.     If the screen is more distant from the GC than 50~pc,
the plotted curves move toward the left.
}
\label{fig:pfraction}
\end{figure}

Broadening increases
the pulse width to $W_{\rm eff} \sim ( W^2 + \tau^2)^{1/2}$ and,
hence, the duty cycle to $\epsilon_{\rm eff} \equiv W_{\rm eff}/P$.
The pulsed fraction becomes
\begin{equation}
\eta_{P}^{\rm (s)} = 
     \eta_P \left\vert\frac{\tilde g(\epsilon_{\rm eff})}  
                           {\tilde g(\epsilon )} \right\vert
 \approx \exp\left[  
          - \left ( \frac{\pi\tau} {2\sqrt{\ln 2} P} \right )^2 
       \right ].
\label{eq:pfraction_eff}
\end{equation}
Figure~\ref{fig:pfraction} shows the pulsed fraction plotted against
frequency for five different pulse periods.  We have assumed that
the GC screen is near Sgr~A${}^*$ ($\deltagc = 50$~pc). 

\begin{figure}[tbh]
\plotfiddle{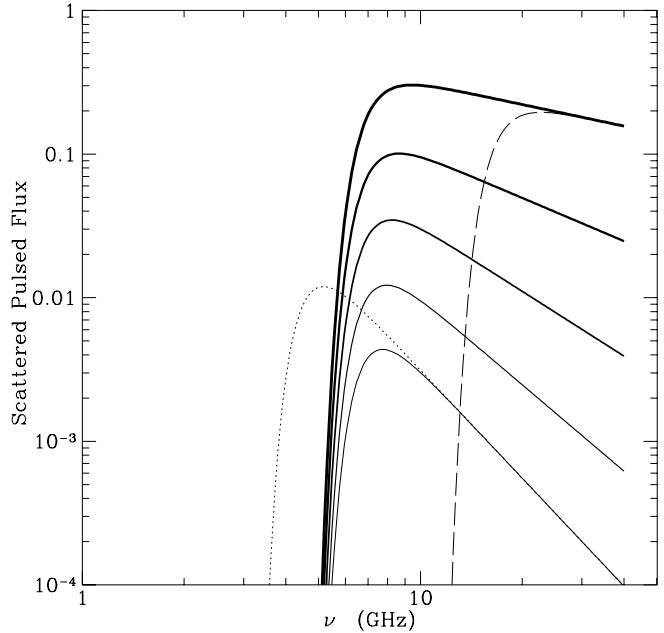}{4.0truein}{0.}{45}{45}{-144}{-52}
\caption[Pulsed Flux]{
The pulsed flux, taking scattering into account, as a function of frequency.
Pulsar spectra of the form $\nu^{-\alpha}$ have been normalized
so that the flux at 1~GHz is the same for all $\alpha$. 
The intrinsic pulse duty cycle is $\epsilon = 0.05$.
{\sl Solid Lines:} $P = 1$~s for $\alpha = 0.5$ (heaviest line),
1, 1.5, 2, 2.5 (thinnest line).  
{\sl Dotted Line:} $P = 5$~s and $\alpha = 2.5$.
{\sl Dashed Line:} $P = 30$~ms and $\alpha = 0.5$.
}
\label{fig:pflux}
\end{figure}

Most pulsars have radio spectra that are power-laws in form, some
objects displaying one or more break points separating power-laws of
different slope (\cite{lor+95}).  Letting the spectrum be $S_{\nu}
\propto \nu^{-\alpha}$, we maximize the {\em pulsed\/} flux density
$\eta_P^{\rm (s)} S_{\nu}$ against frequency to solve for the
corresponding optimal pulse broadening, $\tau_{\rm max} = (\alpha\ln
2/2)^{1/2} P/\pi$.  The frequency at which the pulsed flux density is
maximized can be found from Eq.~\ref{eq:taugc} as
\begin{eqnarray}
\nu_{\rm max} & = & 2.41\, {\rm GHz}\, \left(\frac{f}{\sqrt{\alpha} P}\right)^{1/4} \nonumber \\
 & \sim & 7.3 \, {\rm GHz} \left(\sqrt{\alpha}\Delta_{0.1}P\right)^{-1/4},
\label{eq:critfreq2}
\end{eqnarray}
where $\Delta_{0.1}\equiv \Delta_{\rm GC}/0.1 {\rm kpc}$.
Note that the optimal frequency is a function of pulse period.
At the optimal frequency, the pulsed fraction is
$\eta_p^{(s)} = e^{-\alpha/8}$ and the pulse broadening yields an effective
duty cycle
\begin{equation}
\epsilon_{\rm eff} = \left( \epsilon^2 + \alpha\ln 2/2\pi^2 \right)^{1/2}
\approx 0.19 \sqrt\alpha.
\label{eq:dc_max}
\end{equation}
The approximate equality is for the case where $\epsilon\ll 0.2$
and suggests that, for such pulsars, the search algorithm should
search for no more than $\epsilon_{\rm eff}^{-1} \sim 5$ harmonics
in the Fourier transform. 
Figure~\ref{fig:pflux} shows $\eta_P^{\rm (s)}S_{\nu}$ plotted against
frequency for several pulse periods, $P$, and spectral indices,
$\alpha$.  

There is a barrier of indetectability at low frequencies.
It is obvious that only very luminous, long period pulsars with
shallow spectra are candidates for detection.  The most luminous
pulsars are those that show ``core'' emission (\cite{r90}).  
A good example is
B1933$+$16 with $P = 0.36$~s, distance $D = 7.9$~kpc, and period-averaged
flux densities $S_{400} = 242$~mJy and $S_{1400} = 42$~mJy at 400 and
1400~MHz, respectively (\cite{tml93}).  If placed at the GC, the
pulsed fraction of this pulsar $\ge e^{-1}$ for $\nu \ge 2.4 f^{1/4}$~GHz.
With $\alpha = 1.4$ and $f=170$ (i.e., $\deltagc = 50$~pc), 
we find that $\nu_{\rm max} \approx 10.8$~GHz.
At this frequency, the pulsed flux of B1933$+$16 is 1.8~mJy if placed at the
GC.  Pulsar surveys can reach this flux density through use of large
bandwidths (e.g., 1~GHz).
For pulsars in general, we find that 
the optimal range of search frequencies is
$5\, {\rm GHz} \simless \nu_{\rm max} \simless 18 \, {\rm GHz}$
for
$0.5\simless \alpha \simless 2$ and
$0.1 \simless P \simless 5$~s.

The frequency of maximum, pulsed flux, as we have defined it,
does not correspond to the frequency where S/N is maximized
in a Fourier transform search.  Maximizing S/N
depends on the number of harmonics detectable in the DFT,
and on the variation of system temperature with frequency.
At the optimal frequency, where
approximately $\epsilon_{\rm eff}^{-1} \sim 5$
DFT spikes should be detectable, the S/N can be increased by a factor
$\sim \sqrt 5$. 

We conclude this section with an example search program for
finding GC pulsars in periodicity searches (see Table~\ref{tab:p-dm}).
We consider searches at 5, 8, 10.8 and 15~GHz on a telescope with
sensitivity equal to that for  the National Radio Astronomy Observatory's
Green Bank Telescope, now under
construction.  
The minimum flux density for a single harmonic in the DFT is
\begin{equation}
S_{\rm min} =
    \frac{\eta_T T_{\rm sys}} {G\sqrt{\Delta\nu T}} \sim 130\,\mu {\rm Jy},
\label{eq:smin}
\end{equation}
where $T_{\rm sys}$ is the system temperature, assumed to be 50~K,
$G \sim 2$~K/Jy is the telescope gain, 
the receiver bandwidth is $\Delta\nu = 1$~GHz, 
and the total observing time is $T = $1~hr.  The signal-to-noise threshold for
detection of a signal is assumed to be $\eta_T = 10$.
The 1-GHz bandwidth is substantially larger than that used in most surveys
but is quite feasible at high frequencies.  Moreover, only coarse
channelization is required to combat dispersion smearing. 
For dispersion measures as large as 2000 pc~cm$^{-3}$,
only about 16 channels are needed. 
%
The results in Table~\ref{tab:p-dm} illustrate an analysis 
of the fundamental only; as discussed above, 
analysis of higher harmonics could improve the
quoted sensitivities by a factor of $\sim 2$.  

\begin{deluxetable}{cccccccc}
\tablecaption{Pulsar Detection in Periodicity Searches\label{tab:p-dm}}
\tablehead{
 & & & & \multicolumn{2}{c} {B1933$+$16} & \nl
\cline{5-6} \nl
\colhead{$\nu$} & \colhead{$\tau_{\rm GC}$} 
	& \colhead{$S_{\rm sys}$} & \colhead{$S_{\rm min}$} 
	& \colhead{$\eta_P^{(\rm s)}S_{\nu}$} & \colhead{$(S/N)$} 
	& \colhead{$\tilde P$} & \colhead{$f_L$} \\
\colhead{(GHz)} & \colhead{(s)} 
	& \colhead{(Jy)} & \colhead{($\mu$Jy)} 
	& \colhead{(mJy)} & \colhead{ } 
	& \colhead{(s)}    &  \colhead{ } \\
\colhead{(1)} & \colhead{(2)} 
	& \colhead{(3)} & \colhead{(4)} 
	& \colhead{(5)} & \colhead{(6)}
	& \colhead{(7)} & \colhead{(8)} }

\startdata

\phn5\phd\phn & 1.7\phn\phn & 25 & 130 & 0\phd\phn & $\ll 1$\phn  & 7.8 & $10^{-2.6}$ \nl
\phn8\phd\phn & 0.26\phn    & 25 & 130 & 0.5       & \phn38 & 1.2 & $10^{-2.9}$ \nl
10.8          & 0.08\phn    & 25 & 130 & 1.8       & 138    & 0.4 & $10^{-3.1}$ \nl 
15\phd\phn    & 0.021       & 25 & 130 & 1.6       & 123    & 0.1 & $10^{-3.3}$ \nl
			               
\enddata

\tablecomments{(1) Observation frequency;
	(2) Pulse-broadening time for $\deltagc = 50$~pc;
	(3) $S_{\rm sys} = T_{\rm sys}/G$, where $G$ is the telescope
	    sensitivity in K~Jy${}^{-1}$;
	(4) Minimum detectable flux of fundamental in FFT; 
	(5) Pulsed flux density of PSR~B1933$+$16;
	(6) Signal-to-noise ratio for a search of PSR~B1933$+$16, if it
	    were at the Galactic center;
	(7) Period for which the frequency is the optimum search frequency.
	(8) Fraction of total pulsar population likely to be
	    detectable, assuming nominal spectral index of $-1.4$}
\end{deluxetable}

Two different measures of a search's ability to detect pulsars are 
illustrated in Table~\ref{tab:p-dm}.  The first is
its ability to detect B1933+16, if it were placed in the GC.  Entries in
the Table suggest that large S/N can be achieved at frequencies 
greater than 8 GHz.  Detection at 5 GHz is hopeless.  We also assess the 
fraction of the total GC pulsar population that a search could detect,
which we determine from the luminosity function for radio pulsars.
In the solar neighborhood, for
observation frequencies near 400~MHz, the pulsar luminosity function
has the form $dN/d\log L \propto L^{-\beta}$ with $\beta \approx 1$
(Lyne, Manchester, \& Taylor~1985; \cite{mt77}).  The fraction of
sources with luminosity greater than $L$ is 
$f_L(>L) \approx (1 - L/L_1) L_0/L$, where $L_0, L_1$ are the minimum 
and maximum {\em intrinsic\/} luminosities of a pulsar.  
In the most recent catalog of pulsars
(\cite{tml93}), only one pulsar has a luminosity less than 1
mJy~kpc${}^2$,  so we take $L_0 = 1$ mJy~kpc${}^2$. 
We use a spectral index of $1.4$ to scale in frequency.  
For $S_{\rm min}$ given by Eq.~\ref{eq:smin}, the implied minimum
luminosity at the particular observation frequency is
$\sim 11$ mJy kpc$^2$.  Scaling {\em from\/} the frequencies in
the Table then yields the equivalent $L$ at 400~MHz and the fraction of objects
that exceed this minimum.  We also tabulate the period, $\tilde P$,
at which the tabulated frequency is, in fact, the optimal frequency,
as given by Eq.~\ref{eq:critfreq2}.  As can be seen,  this period
exceeds that of any known radio pulsar at 5~GHz, but decreases progressively
to $\sim 0.1$~s at 15~GHz.   The fraction of objects that exceed 
the minimum luminosity also decreases in going from 5 to~15~GHz.  A key disadvantage to using higher frequencies is that, all other
things being equal, the telescope time needed to search a region of
fixed angular size will scale as 
the solid angle of the telescope beam,  $\propto\nu^{-2}$.   This amounts
to nearly an order of magnitude increase in time needed in going from
5 to~15~GHz.

The results in Table ~\ref{tab:p-dm} are only illustrative.  Pulsars
show a wide range of spectral indices and there are possible correlations
between period and luminosity that we have ignored. 
Also, we have been assuming a constant pulse-broadening value for
pulsars in the GC.  In actuality, the spatial distribution of pulsars
in the GC will create a range of pulse broadening.  The extreme values
of pulse broadening we have estimated above will be obtained if the
angular distribution of pulsars is concentrated toward Sgr~A${}^*$,
yet are still quite far from the scattering screen.  However, pulsars
spread throughout the region will show a wide range of broadening.
Objects close to the screen will show small scattering, asymptotic to
the TC value as $\Delta_{\rm psr} \to 0$.  Periodicity searches at
different (high) frequencies will therefore sample the population at
different $\Delta_{\rm psr}$.

\section{SEARCHES FOR PULSARS IN IMAGING SURVEYS FOR POINT SOURCES}
\label{sec:imaging}

Scattering preserves the total flux density while attenuating the
pulsed flux.  Pulsars may therefore be searched for as objects with
pulsar-like spectra and polarization in aperture synthesis surveys of
the GC.  Angular broadening dilutes the surface brightness, but we
show here that this dilution denigrates a synthesis survey far less
than pulse broadening does a periodicity search.  Indeed Sgr~A${}^*$,
the OH/IR stars (\S\ref{sec:intro}), and the GC transient
(\cite{z+92}) demonstrate that heavily scattered, yet sufficiently
luminous, sources in the GC can be detected with aperture synthesis
instruments.

Table~\ref{tab:image} illustrates detection of pulsars in an imaging
survey with the VLA.  We assume observations are conducted with a
total integration time of 1~hr and total bandwidth
$\Delta\nu = 50$~MHz at 1.4 and 5~GHz and $\Delta\nu=3$ MHz at 0.33~GHz.  
As before, we consider the detection of both
B1933+16 and a fraction of the total pulsar population
(\S\ref{sec:detection}).  
Recent observations at 0.33~GHz (e.g., \cite{fkgc95}) have demonstrated the
capability of making low-frequency observations with the VLA which
approach the thermal noise limit in the A-configuration.  Within Sgr~A
West, i.e., within 5~arc~min of Sgr~A${}^*$, strong thermal absorption
is seen at 0.33~GHz.  This absorping gas strongly attenuates
Sgr~A${}^*$ below 1~GHz (\cite{apeg91}; \cite{paeggsz89}) and may
contribute to the extreme scattering of Sgr~A${}^*$ at higher
frequencies.  However, heavily scattered OH masers are seen up to
25~arc~min from Sgr~A${}^*$ (\cite{fdcl94}).  Over this larger region,
the free-free optical depth is only $\sim 1$.  Therefore, absorption
may attenuate source fluxes but should not render the entire
scattering region opaque.


\begin{deluxetable}{cccccc}
\tablecaption{Pulsar Detection in an Imaging Survey\label{tab:image}}
\tablehead{ & & & \multicolumn{2}{c} {B1933$+$16} & \\
\cline{4-5} \\
\colhead{$\nu$} & \colhead{$\theta_{\rm GC}$} 
	& \colhead{$I_{\rm min}$} 
	& \colhead{$I$} & \colhead{$(S/N)$} 
	& \colhead{$f_L$} \\
\colhead{(GHz)} & \colhead{(\arcsec)} 
	& \colhead{(mJy/beam)} 
	& \colhead{(mJy/beam)} & \colhead{ } 
	& \colhead{ } \\
\colhead{(1)} & \colhead{(2)} 
	& \colhead{(3)} 
	& \colhead{(4)} & \colhead{(5)}
	& \colhead{(6)} }

\startdata

0.3       & 11\phd\phn\phn & 2.1\tablenotemark{a}\phn & 40\phd\phn & \phn19 & 0.007 \nl
1.4       & 0.62           & 0.17                     & 33\phd\phn & 194    & 0.012 \nl
5\phd\phn & 0.05           & 0.13                     & \phn6.7    & \phn52 & 0.002 \nl
			               
\enddata
\tablecomments{(1) Observation frequency;
	(2) Scattering angle for Galactic center source (i.e., Sgr~A${}^*$);
	(3) Minimum detectable brightness;
	(4) Brightness of PSR~B1933$+$16, if it were at the Galactic center;
	(6) Signal-to-noise ratio for a search of PSR~B1933$+$16, if it
	    were at the Galactic center;
	(7) Fraction of total pulsar population likely to be
	    detectable, assuming nominal spectral index of $-1.4$}

\tablenotetext{a}{Assumes bandwidth at 0.33~GHz of $\Delta\nu = 3$~MHz.}

\end{deluxetable}

For the Table, we have assumed that pulsar scattering diameters 
are the same as that of Sgr~A${}^*$.  From considerations in
\S\ref{sec:geometry}, the scattering diameter of a pulsar at distance
$D> \dgc-\deltagc$ is
\begin{equation}
\theta_{\rm psr}(D) = \left (1 + \frac{D-\dgc}{\deltagc} \right )
                  \left (\frac{\dgc}{D} \right)
                  \theta_{\rm GC}.
\label{eq:theta_psr}
\end{equation}
Pulsars closer to the screen than Sgr~A${}^*$ will show less angular
broadening than Sgr~A${}^*$ while a pulsar only $\deltagc$ further
than Sgr~A${}^*$ will be twice the diameter.  An object 0.5~kpc beyond
Sgr~A${}^*$ will be 10 times the diameter (for $\deltagc = 50$~pc), or
$13\arcsec\,\nu_{\rm GHz}^{-2}$.  The angular diameters of pulsars in
the vicinity of Sgr~A${}^*$ are therefore well matched to the
synthesized beams of the VLA in its four configurations, though only
the larger configurations filter out the intense extended emission
sufficiently.  Note that the pulsar B1933+16 is easily detectable if
placed in the GC at 1.4 as well as 5~GHz.  Moreover, about 1.2\% of
all pulsars will be detectable at 1.4~GHz (for an assumed spectral
index of 1.4), about 10 times larger than the fraction found in a
periodicity search.

\section{GALACTIC CENTER PULSAR POPULATIONS}\label{sec:population}

The preceding sections have assumed the existence of pulsars in the
GC.  In this section we assess how many pulsars are likely to exist in
the GC and use the results of \S\S\ref{sec:detection}
and~\ref{sec:imaging} to indicate how many of these are in fact
detectable.  The possibility of a large number of stellar remnants at
the GC has been discussed by a number of other authors, though their
focus has been on the central few parsecs (e.g., Saha, Bicknell, \&
McGregor~1996) and on either black holes (\cite{m93}) or white dwarfs
(\cite{hrrtcm96}). Our focus will be on neutron stars exclusively and,
based on the size of the scattering region, will be on a larger size
scale than considered previously.  We estimate the number of GC
pulsars by considering both the integrated star formation history of
the GC and the effects of a recent star burst.  Finally, we also
discuss the implications of the detection of GC pulsars on our
understanding of the dynamics and star formation history of the GC.

We first consider the number of pulsars likely to have been formed if
the star formation in the GC has been occurring at a constant rate
over the Galaxy's history.  In actuality, factors such as episodic
star bursts, an initial mass function (IMF) different from that in
the disk, and the shape of the central potential and its effects on
gas dynamics may all have contributed to a variable star formation
rate.  The assumption of a constant star formation rate is nonetheless
useful for illustration and comparison with the starburst model.

\subsection{Steady Star Formation}

We extrapolate from the current mass of the GC, for an assumed IMF, in
order to estimate the number of pulsars likely to have formed within
the GC.  Various methods, summarized by Genzel, Hollenbach, \&
Townes~(1994), constrain the mass enclosed within the central 100~pc
to be in the range $\sim 4 \times 10^8$--$10^9$~M${}_{\sun}$.  An
estimate of the enclosed mass derived from the stellar luminosity
distribution, and assuming a constant mass-to-light ratio, is
approximately $7 \times 10^8$~M${}_{\sun}$, which we adopt as a
nominal value for the mass interior to 100~pc.  Most of this mass is
in the form of stars (\cite{ght94}).  Because we do not know the form
of the IMF in the GC, we considered different IMFs, one appropriate
for the Galactic disk (\cite{ms79}) and one that accounts for a
stronger weighting of the IMF toward high-mass stars during starbursts
(\cite{rltlt80}).  For these IMFs, and different assumed ranges of
masses for main-sequence stars that result in neutron stars, we find
that 4--$12\times 10^7$ neutron stars have formed in the GC.

Implicit in this extrapolation is the assumption that star formation
and evolution in the GC is similar to that in the disk.  Morris~(1993)
has discussed the conditions in the central 1--10~pc and how they
could both inhibit massive star formation and alter the evolution of
those stars which do form.  Meynet et al.~(1994) have shown that
massive, high-metallicity stars may lose a substantial fraction of
their mass during their post--main-sequence evolution and they suggest
the remnants of such stars would be white dwarfs rather than neutron
stars (and black holes).  Even if such processes are operative,
however, the presence of some massive stars is clearly indicated.
There are a number of \ion{H}{2} regions within 15\arcmin\ of
Sgr~A${}^*$ which appear to be powered by OB and/or WR stars (e.g.,
Figer, McLean, \& Morris~1996; \cite{pggmntt96}); observations of
H${}_2$ ($v = 1 \to 0$) S(1) emission in the inner 2\arcdeg\ imply
excitation by a UV radiation field (Pak, Jaffe, \& Keller~1996); and
star counts within the inner 2\arcdeg\ reveal a number of stars with
dereddened absolute magnitudes of $M_K \lesssim -9$, suggestive of
stars with $M \gtrsim 5$~M${}_{\sun}$ (Catchpole, Whitelock, \&
Glass~1990; \cite{bbcfn94}).  Bearing these caveats in mind, we shall
use our estimate of $10^7$--$10^8$ neutron stars in the GC.

The actual number of active {\em and detectable\/} pulsars in the GC will be substantially smaller.
Firstly, the mean space velocity of pulsars is 500 km~s${}^{-1}$
(\cite{ll94}).  A considerable fraction, $\gtrsim 0.25$, is not bound
to the Galaxy.
Only those pulsars forming the lowest-velocity tail, $\lesssim 100$
km~s$^{-1}$ or about 16\% of the objects, are bound to the
central 100~pc.  Secondly, the lifetime for pulsar radio emission is
$\sim 10^7$~yr for strong field objects with surface fields
$\sim 10^{12}$~Gauss.  If the Galactic bulge is $\sim 10^{10}$~yr old, only
a fraction $\sim 10^{-3}$ of the neutron stars formed in the central
100~pc will still be active pulsars.  Hence, of the $\sim
10^7$--$10^8$ neutron stars formed in the GC over the lifetime of the
Galaxy, $\sim 10^3$--$10^4$ are likely to be active pulsars still
within the central 100~pc.  Of these, only a fraction $f_{\rm b} \sim
0.2$ will be beamed toward us (\cite{b90}; \cite{lm88}; \cite{nv83}).
If about 1\% of these can be detected, as in an imaging survey using
the VLA, then only a few to $\sim 20$ objects are expected to be
found.  A periodicity search that finds only the 0.1\% most luminous
pulsars is unlikely to find any pulsars from this population.

One effect which would increase this number is the production of
recycled pulsars in binaries resulting from tidal captures and stellar
collisions.  The central stellar density may exceed
$10^7$~pc${}^{-3}$, sufficiently high that stellar encounters should
be frequent and that processes for recycling pulsars should be
operative.  Fabian, Pringle, \& Rees~(1975) estimate the tidal capture
rate (see also \cite{hut+92}).  If the number of neutron stars formed
in the GC has been $\sim 10^7$--$10^8$, then $\sim 10^3$ tidal
captures or collisions will have occurred over the Galaxy's history.
Not all of the binaries so formed will produce a recycled pulsar and
primordial binary systems could also contribute to the population of
recycled pulsars.  Regardless of formation mechanism, the detection of
a large number of soft, compact X-ray sources toward the GC (e.g.,
\cite{pt94}) supports the possibility of an enhanced number, relative
to the disk, of neutron-star binary systems in the GC.  Although some
of these X-ray sources are probably black-hole systems, the fraction
of millisecond pulsars in the total pulsar population in the GC could
nevertheless be enhanced substantially relative to a value $\sim 0.5$
in the disk (\cite{bl95}; \cite{lbdh93}).

Unfortunately, the fastest spun-up pulsars will not be detectable in a
periodicity search, though they may appear in an imaging search.
Modestly spun-up pulsars with lifetimes 10--100 times longer than
those of most pulsars may be the best targets.  In this respect it is
interesting that 15\% of the spun-up pulsars in a recent listing
had periods greater than 100~ms (\cite{k95}), though none of these
objects would themselves be detectable if placed at the GC.

\subsection{Star Bursts}
The discussion above assumes a constant star formation rate within the
GC.  A recent starburst has been invoked (e.g., \cite{har95};
\cite{ght94}; \cite{sof94}; but see also \cite{m93}) to explain one or
more features of the GC.  Estimates for the age of the burst are
$10^6$--$10^7$~Myr with the total number of neutron stars formed being
$10^2$--$10^6$.  Even if a burst per~se has not occurred, several
lines of evidence point to an enhanced star formation rate, relative
to the disk rate, during the recent past in the GC.  These include the
density of supernova remnants within the central 10\arcdeg\, which is double
that found in the disk (\cite{g94}), and the presence of  
a large number of shell and arc structures in molecular gas, also 
suggestive of recent supernova activity (\cite{hosthm96}).

The estimated range of ages for the most recent starburst is
sufficiently small that all of the pulsars formed during the burst
should still be active.  High velocity pulsars will escape the GC in a
time short compared to the burst.  Indeed, Hartmann~(1995) advocates
searches for these escaped pulsars as a means for diagnosing the
recent star formation history in the GC.  Depending upon the number of
neutron stars created in the burst, the total number of pulsars in the
GC from the low-velocity tail of the pulsar velocity distribution
could number $10$--$10^5$.  That is, the number of active pulsars from
the burst population could exceed that from a constant star formation
rate over the history of the GC.

The number of detectable pulsars will be smaller than the total number
active and within the GC due to selection effects.  In
\S\S\ref{sec:detection} and \ref{sec:imaging} we estimated the
fraction of pulsars sufficiently luminous to be detected, $f_L$, based
on the local pulsar luminosity function.  This fraction is
$10^{-2}$--$10^{-5}$, depending upon the search method and frequency.
Taking into account the non-isotropic nature of pulsar emission, the
resulting fraction of detectable pulsars is $\sim 10^{-3}$--$10^{-6}$.
Thus, the number of active, detectable pulsars from the central 100~pc
of the Galaxy could be $\lesssim 1$ to as many as 100.  A constant
star formation rate over the GC's history will likely contribute
$\lesssim 10$ pulsars to the total number detectable.

\subsection{Pulsars as Probes of the Galactic Center}

We conclude this section with a brief discussion of the utility of
pulsars for constraining the dynamics and star formation history of
the GC.  The number of pulsars detected, together with a careful
accounting for selection effects, can be used to constrain the past
star formation rate of the GC.  If any pulsars can be detected as
sources of pulsed emission, the distribution of spin periods; their
period derivatives, $\dot P$; and dispersion measures, DM, and pulse
broadening should also be measurable.  The distribution of periods can
be used in estimating the past star formation rate (e.g.,
\cite{lbdh93}).  Measurement of $\dot P$ has the potential to
constrain the GC gravitational potential, much like the determination
of a negative $\dot P$ for B2127$+$11 has set limits on its
acceleration from the potential of the globular cluster M15
(\cite{wkmbfd89}).  As noted in \S\ref{sec:intro}, the most recent
model for the Galactic distribution of ionized gas has few constraints
toward the inner Galaxy.  Even a few DM and pulse broadening
measurements would prove useful for improving substantially our
knowledge of conditions there.  Moreover, gas velocities, particularly
near Sgr~A${}^*$, are large, in some cases exceeding 100 km~s${}^{-1}$
(Schwarz, Bregman, \& van~Gorkom~1989).  DM monitoring programs can
expect to detect changes in the DM on time scales of a month, which in
turn could be used to probe the AU-scale structure of ionized gas in
the GC.  Even if pulsars can only be detected in an imaging survey,
their angular diameters would still provide both constraints on the
distribution and a probe of the small-scale structure of ionized gas
in the GC.

Finally, the clock-like nature of pulsars has prompted speculation
about their utility for studying the constituents and mass
distribution in the GC.  Wex, Gil, \& Sendyk~(1996) and, earlier,
Paczynski \& Trimble~(1979) have discussed the microlensing and
gravitational time delay of GC pulsars.  The time delay, in
particular, can be used to constrain the mass of any central object.
Such determinations, however, will be possible at only the highest
frequencies where pulsed emission is manifested.

\section{CONCLUSIONS AND RECOMMENDATIONS}\label{sec:conclude}

The extreme level of scattering seen toward the Galactic center
(1.3\arcsec\ of broadening for Sgr~A${}^*$ at 1~GHz) produces
considerable pulse broadening.  The amount of pulse broadening depends
upon the Galactic center-scattering screen distance, $\deltagc$, but
at minimum, the pulse broadening is 6.3~{\em
seconds}.  If the screen is closer to the Galactic center, the pulse
broadening is increased.  Radio pulsars within about
15\arcmin--1\arcdeg\ of Sgr~A${}^*$ and behind the GC scattering
screen will only be detected by periodicity searches conducted at
frequencies much higher than used heretofore {\em or\/} by imaging surveys.

Periodicity searches at 5~GHz will be capable of detecting pulsars
with canonical spin periods $\sim 1$~s, though with much reduced
sensitivity if $\deltagc \ll \dgc$.  At 10~GHz, pulse broadening of
millisecond pulsars is a significant fraction of a period even if
$\deltagc \approx \dgc$.  If $\deltagc \sim 100$~pc, pulsars with
periods shorter than $0.5$~s at 5~GHz and 34~ms at 10~GHz will be
inaccessible.  To reach pulsars with the Crab pulsar's period (33~ms)
such that a 10\% duty cycle pulse may be seen without broadening, the
observing frequency must be at least 18~GHz.  While the Crab pulsar
itself is undetectable at this frequency, other young pulsars with
shallower spectra (e.g., \cite{lor+95}) may be strong enough.  Such
objects are rare, so we conclude that imaging surveys at fairly low
frequencies (e.g., 1.4~GHz) are much more likely to be successful in
finding Galactic center pulsars.

The optimal frequency for conducting any periodicity searches is that
which maximizes the pulsed flux density and is given by 
$\nu_{\rm max} = 
    7.3 \,{\rm GHz}\left(\sqrt{\alpha}\Delta_{0.1}P\right)^{1/4}$
(Eq.~\ref{eq:critfreq2}).  The effective duty cycle of a scattered
pulsar at the optimal frequency is $\approx 0.2\sqrt{\alpha}$ and
suggests that search algorithms need not consider more than
$\epsilon_{\rm eff}^{-1} \sim 5$ harmonics in the DFT.

The angular broadening for a pulsar at the GC will be similar to that
of Sgr~A${}^*$; the broadening for a pulsar more distant than
Sgr~A${}^*$ but seen through the same scattering screen is given by
Eq.~\ref{eq:theta_psr}.  These angular sizes are well matched to the
synthesized beams of the VLA in its four configurations.

Finally, the number of pulsars detectable in the Galactic center
depends upon the star formation history of the Galactic center.  The
largest number of detectable pulsars will occur if the GC has
undergone a recent episode ($\lesssim 10^7$~yr) of massive star
formation.  The number of active, detectable pulsars in the GC may be
as many as 100.  Smaller starbursts and the star formation over the
Galaxy's history can contribute another 1--10 pulsars.

\acknowledgements
We thank Z.~Arzoumanian, D.~Chernoff, D.~Hartmann,  
and M.~McLaughlin for discussions; P.~Piper for assistance with Fig.~\ref{fig:geometry}; and the referee, D.~Lorimer, for useful suggestions.
This research was supported by the National Science Foundation
through grant AST~92-18075 to Cornell University. 


\end{document}